`\pdfoutput=1`
\documentclass[aps,prl,floatfix,twocolumn,showpacs,preprintnumbers,amsmath,amssymb,a4paper,superscriptaddress]{revtex4-1}
\usepackage{braket}
\usepackage{dsfont}
\usepackage{amsmath}
\usepackage{amssymb}
\usepackage{amsfonts}
\usepackage{mathtools}
\usepackage{graphicx}
\usepackage{dcolumn}
\usepackage{bm}
\usepackage{multirow}
\usepackage{leftidx}
\usepackage{color}
\usepackage{mathtools}
\usepackage{MnSymbol}
\usepackage[mathscr]{eucal}
\usepackage[german,english]{babel}

\begin{document}

\title{Non-adiabatic molecular association in thermal gases driven by radio-frequency pulses}

 \author{P. Giannakeas}
 \email{pgiannak@pks.mpg.de}
 \affiliation{Max-Planck-Institut f\"ur Physik komplexer Systeme, N\"othnitzer Strasse 38, 01187 Dresden, Germany}

 \author{L. Khaykovich}
 \email{lev.khaykovich@biu.ac.il}
 \affiliation{Department of Physics, QUEST Center and Institute of Nanotechnology and Advanced Materials, Bar-Ilan University, Ramat-Gan 5290002, Israel}

 \author{Jan-Michael Rost}
 \email{rost@pks.mpg.de}
 \affiliation{Max-Planck-Institut f\"ur Physik komplexer Systeme, N\"othnitzer Strasse 38, 01187 Dresden, Germany}

 \author{Chris H. Greene}
 \email{chgreene@purdue.edu}
 \affiliation{Department of Physics and Astronomy, Purdue University, West Lafayette, Indiana 47907, USA}
 \affiliation{Purdue Quantum Science and Engineering Institute, Purdue University, West Lafayette, Indiana 47907, USA}

 \date{\today}

\begin{abstract} The molecular association process in a thermal gas of $^{85}$Rb is investigated where the effects of the envelope of the  radio-frequency field are taken into account.
	For experimentally relevant parameters our analysis shows that with  increasing pulse length  the corresponding molecular conversion efficiency exhibits low-frequency interference fringes
 which are robust under thermal averaging over a wide range of temperatures.
This  dynamical interference phenomenon is attributed to St\"uckelberg phase accumulation between the low-energy continuum states and the dressed molecular state which exhibits a  shift proportional to the envelope of the radio-frequency  pulse intensity.
\end{abstract}


\maketitle
External fields are widely used in order to probe, tune and control various aspects of atomic matter.
For example, in the field of ultracold atomic physics dc magnetic fields constitute the main experimental means for the creation and manipulation of molecules via Feshbach resonances \cite{chin_feshbach_2010,stwalley1976prl,tiesinga1993pra}.
Techniques involving radio-frequency (RF) magnetic fields are of exceptional importance since they are highly adjustable in experiments \cite{kaufman2009pra,tscherbul2010pra}.
Indeed, the additional magnetic RF field modulation enables the investigation of cold molecule formation \cite{hanna2007pra,thompson2005prl,yijue2016chemphys} or heteronuclear association/dissociation processes in a microgravity environment \cite{dincao2017prl}, association of Efimov trimers \cite{nakajima2011prl,machtey2012prl,lompe_radio-frequency_2010,tscherbul2011pra} or manipulation of Feshbach collisions \cite{hudson2015prl,sykes2017pra,yijue2017pra}.
Beyond cold physics, external fields are also used in ultrafast physics where short laser pulses probe photoionization processes \cite{ivki+06}.
In such systems the pulse envelope plays a crucial role since it induces a time-dependent AC-Stark shift of the energy levels of the system whereas the light-pulse derivatives yield a dynamic interference in photoionization cross-sections \cite{tosa+15,baghery2017prl,ning2018prl,toyota2007pra,toyota2008pra,demekhin2012prl,deho+13,deho+17}, \cite{com2}.

In this letter, the RF-induced association process in an ultracold thermal gas is investigated and it is shown that the RF field envelope plays a crucial role.
This allows us to extend the concept of dynamical interference in strong field ionization into the realm of ultracold physics and to explore its impact on the production of cold molecules.
In Ref.\cite{thompson2005prl} experimental evidences suggested that RF association in a thermal gas can exhibit Rabi-like oscillations in the molecular conversion efficiency (MCE) as a function of the duration of the RF field.
On the other hand, the corresponding theoretical studies in Ref.\cite{hanna2007pra} show that in this range of parameters any coherence in the MCE is completely smeared out by thermal averaging.
Evidently, these studies still pose an intriguing question for the RF association processes in thermal gases: Under what conditions does the RF molecule formation display interference fringes that survive thermal averaging?
Our study tackles this question: a gas of $^{85}$Rb  atoms with density $n=10^{11}/$cm$^{3}$ is considered for which the gas temperature varies from $T=20\,$nK  up to $T=50\,$nK , and for an RF field driving frequency that can associate continuum states near the dissociation threshold.
In this temperature range and density, our analysis predicts that a {\it strong pulsed} RF field induces  dynamic interference, which remains robustly observable in the MCE even after thermal averaging. 
In the limit of weak pulses a monotonic increase of the MCE with pulse length is found, consistent with previous theory \cite{hanna2007pra}.

Our prototype two-body system consists of $^{85}$Rb  atoms in the presence of a broad Feshbach resonance located at $B_0=155\,G$.
The corresponding two-body collisions can be modeled by an effective single channel Hamiltonian where the low collision energies involve $s$-waves only.
In the center-of-mass frame, the two cold atoms experience an RF pulse that can associate them into a molecular state.
This process is addressed by the following time-dependent model Hamiltonian in the relative degrees of freedom,

\begin{equation}
	H(r,t)={\cal T}-V_0 \theta(r_0-r)+\eta (r,t) \cos( \omega t),
	\label{eq:1}
\end{equation}
where $\theta (\cdot)$ is the step function and {$\cal T$} represents the two-body kinetic energy operator. 
The s-wave interactions are modeled via a spherical well of depth $-V_0$ (with $V_0>0$) and range $r_0$. 
The third term of $H(r,t)$ refers to the RF pulse with an angular frequency $\omega$ and pulse envelope $\eta(r,t)=\eta_0 \theta(r_0-r)\chi(t)$, where the explicit time-dependent factor $\chi(t)$ reads
\begin{equation}
	\chi(t) =
	\begin{cases}
		\sin^2[\frac{\pi}{2}(\frac{t}{\tau_0})]& 0\leq t < \tau_0\\
				\hfil 1 & \tau_0 \leq t < \tau_\mathrm{c}+\tau_0\\
				\sin^2[\frac{\pi}{2}(\frac{t-\tau_\mathrm{c}}{\tau_0})]& \tau_\mathrm{c}+\tau_0 \leq t \leq \tau_\mathrm{c}+2\tau_0.
 \end{cases}
	\label{eq:2}
\end{equation}
Here $\eta_0$ is the strength of the pulse, $\tau_0$ indicates the turn on/off time of the pulse and $\tau_\mathrm{c}$ refers to the time interval when the pulse strength is constant.

The adopted field-free Hamiltonian possesses only one bound state near the threshold, with energy $E_b/h=10.108\,$kHz [for details see \footnote{The spherical well depth is $V_0/h=15.872\,$MHz whereas its range is $r_0=82.3~a_0$, equalling the van der Waals length scale for two $^{85}$Rb atoms yielding a scattering length $a_s=2083.9~a_0$}].
For the pulse we consider a driving frequency of the RF field at $\omega/2\pi=10.11\,$kHz which resonantly couples the molecular state with the near-threshold continuum states.

\begin{figure}[t]
\centering
 \includegraphics[scale=0.43]{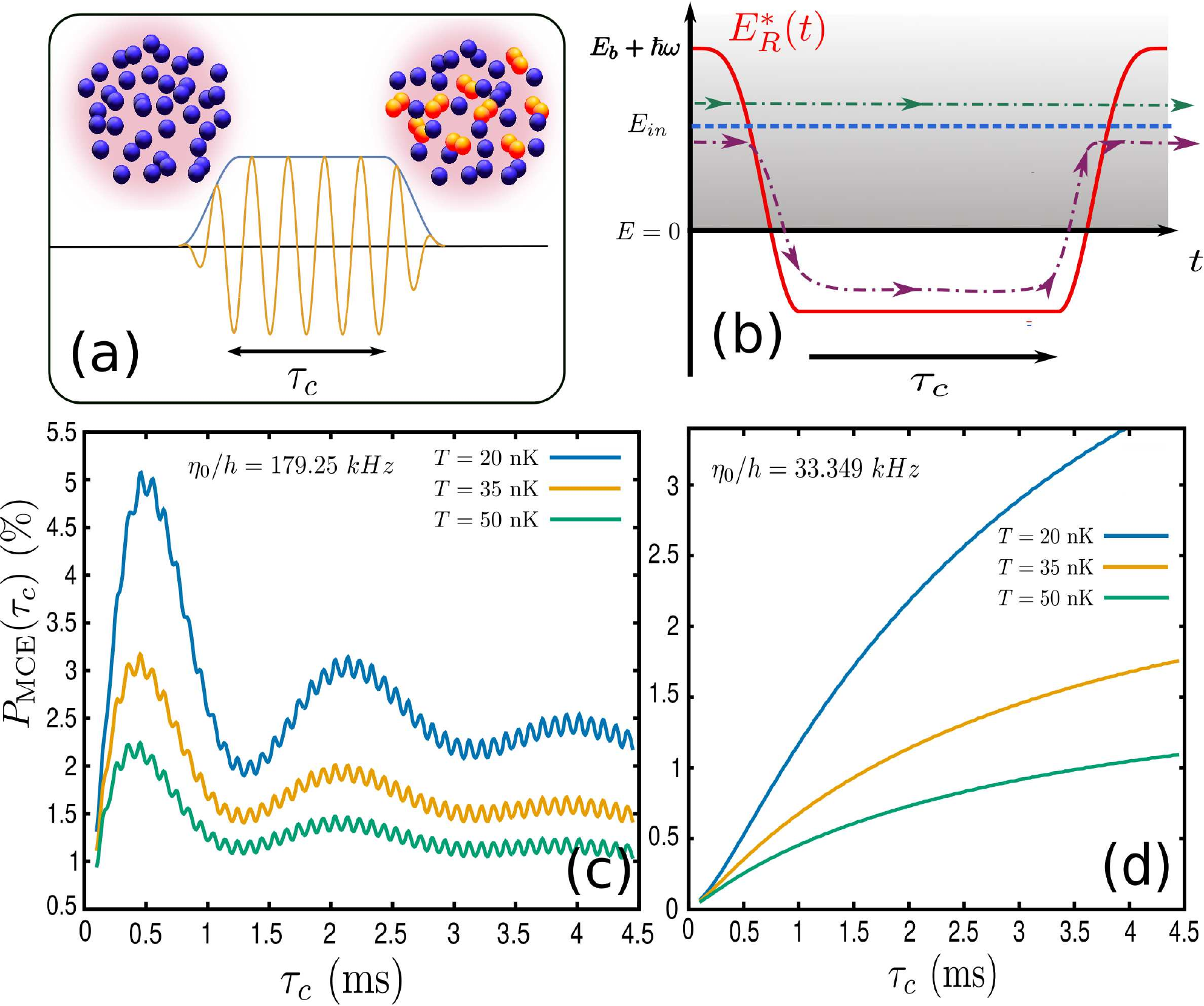}
 \caption{(Color online) (a) A schematic figure shows a thermal $^{85}$Rb gas exhibiting molecular association after the RF pulse, where the blue line indicates the envelope of the pulse with plateau time $\tau_\mathrm{c}$.
	 (b) An illustration of the time evolution of the diabatic energies during the pulse in the rotating-wave approximation, where $E_\mathrm{R}^*(t)$ (red solid line) refers to the shifted dressed bound state energy that saturates to $E_b+\hbar\omega$ at the beginning and end of the pulse. $E_\mathrm{in}$ (blue dashed line) represents the energy of a continuum state.
 The green and purple dashed-doted lines depict the two different pathways that interfere at the end of the pulse.
 (c) and (d) $P_\mathrm{MCE}(\tau_\mathrm{c})$ from Eq.\eqref{eq:4} as a function of the plateau time $\tau_\mathrm{c}$, for three different temperatures, $20\,$nK  (blue), $35\,$nK  (orange) and $50\,$nK  (green), where the maximum of the pulse envelope is $\eta_0/h=179.25\,$kHz  and $\eta_0/h=33.349\,$kHz , respectively.
 The density of the thermal gas is $n=10^{11}/$cm$^{3})$. 
 }
\label{fig1}
\end{figure}
Fig.\ref{fig1}(a) illustrates the initial thermal gas of $^{85}$Rb atoms exposed to an RF pulse of duration $\tau_\mathrm{c}$, after which a fraction of the gas is converted into molecules.
 Modeled with the Hamiltonian shown in Eq.(\ref{eq:1}) the two particles are initially in a continuum state and the pulse induces free-to-bound and free-to-free transitions.
 The time-dependent wavefunction is expanded into field-free Hamiltonian eigenstates, namely $\ket{\psi(t)}=\sumint_\alpha \ket{\alpha}e^{-i E_\alpha t}C_\alpha(t)$, where the integration (summation) runs over {\it energy-normalized} continuum (bound) states which asymptotically obey standing wave boundary conditions.
Note that $\alpha$ represents a collective index containing the relevant quantum numbers.
This basis converts the time-dependent Schr\"odinger equation into a set of first order coupled equations
\begin{eqnarray}
	i\hbar \partial_t C_b(t)&=& \Gamma_{bb}(t) C_b(t) + \int\!\!\! d\epsilon\, \Gamma_{b \epsilon}(t) e^{i(E_b-\epsilon)t/\hbar} C_\epsilon(t) \cr
	i\hbar \partial_t C_\epsilon(t)&=& \Gamma_{\epsilon b}(t) e^{-i(E_b-\epsilon)t/\hbar}C_b(t) + \int\!\!\! d\epsilon '  \Gamma_{\epsilon \epsilon '}(t) e^{i(\epsilon -\epsilon ')t/\hbar} C_{\epsilon '}(t),
	\label{eq:3}
\end{eqnarray}
where $\Gamma_{ij}(t) = \cos(\omega t)\braket{i|\eta(r,t)|j}$ and there is no summation over bound states  since the Hamiltonian in Eq.(\ref{eq:1}) has only one.
Eq.(\ref{eq:3}) is numerically solved following Ref.\cite{tarana2012pra}, with a box-state discretization of the continuum [for details see \footnote{The box size is given by $R_{\rm{box}}>p_{max} t_{\rm{f}}/\mu$, where $\mu$ is the reduced mass, $p_{\rm{max}}$ is the momentum of the highest continuum state (hcs) and $t_{\rm f}$ is the time at the end of the pulse. This ensures that propagating the hcs will not reach the boundary of the box by the end of the pulse. In the calculations presented in the main text the hcs has energy $E_{\rm max}=15.5 \hbar \omega$ yielding an $R_{\rm box}=1.8~10^{7}~\rm{a.u}$}].

Due to the thermal energy distribution, there is no preferred initial continuum state, and therefore the transition probability density $|C_b(t)|^2$ to occupy the molecular state must be thermally averaged over a Maxwell-Boltzmann (MB) distribution of the initial continuum state energies to compute the fraction of atoms converted into molecules.  Calculated after the pulse at time $t_\mathrm{f}=\tau_\mathrm{c}+2\tau_0$ the MCE reads as a function of plateau length $\tau_\mathrm{c}$ \cite{hanna2007pra}:
\begin{equation}
	P_\mathrm{MCE}(\tau_\mathrm{c})\equiv\frac{2N_m}{N}=2 n \lambda_T^3 \int_0^\infty dE_\mathrm{in}e^{-\frac{E_\mathrm{in}}{k_B T}} |C_b(t_\mathrm{f})|^2,
	\label{eq:4}
\end{equation}
with Boltzmann constant $k_B$,  temperature $T$,  density $n$,  and the thermal de Broglie wave length $\lambda_T=\sqrt{2\pi \hbar/(\pi m_\mathrm{Rb}k_B T)}$.
Note that Eq.(\ref{eq:4}) is valid for $N_m<N/2$.
In the numerical solution of the TDSE, a box discretized continuum state is chosen as an initial state and at the end of the pulse the corresponding box-normalized transition amplitude to the bound state is rescaled by the energy-normalization constant. The energy-normalized $|C_b(t_{\rm f})|^2$ is then thermally averaged for an ensemble of 2800 different initial states sampling an energy interval $E_{in}=[0,~0.97\hbar \omega]$.

For a thermal gas of $^{85}$Rb atoms with density $n=10^{11}/$cm$^{3}$, Figs.\ref{fig1}(c) and (d) depict the numerically calculated $P_\mathrm{MCE}(\tau_\mathrm{c})$ for different pulse plateau times $\tau_\mathrm{c}$ and for three different temperatures: $T=20$ nK (blue line), $T=35\,$nK  (orange line) and $T=50\,$nK  (green line).
The peak of pulse envelope is  $\eta_0/h=179.25\,$kHz  in Fig.\ref{fig1}(c) and $\eta_0/h=33.349\,$kHz  in Fig.\ref{fig1}(d) and the ramp on/off time of the pulse is $\tau_0=0.65~2 \pi/\omega$ (with $\omega/2 \pi=10.11\,$kHz) for both panels.
Note that this particular choice of $\tau_0$ is comparable to the response time scale $\tau_b=\hbar/E_b$ of the system and ensures that the resulting dynamics is in the non-adiabatic regime.
Figs.\ref{fig1}(c) and (d) demonstrate striking qualitative differences in the MCE as the pulse strength $\eta_0$ varies.
For strong pulses, Fig.\ref{fig1}(c) shows that as $\tau_\mathrm{c}$ increases the molecular formation displays two types of oscillatory behavior: a fast one which has a frequency equal to $\omega/2\pi$ and a slow one with frequency $\nu=0.532\,$kHz.
We observe that the interference fringes survive the thermal averaging process which would usually tend to smear out  coherence features.
Indeed, the  MCE for  a weak pulse (Fig.\ref{fig1}(d)) does not exhibit any interference,  analogous to the observations in Ref.\cite{hanna2007pra}.
Moreover, for weak pulses, Fig.\ref{fig1}(d), the molecular conversion increases monotonically with the pulse duration, whereas in the case of a strong pulse, Fig.\ref{fig1}(c), the molecule formation probability saturates already in the same range of times $\tau_\mathrm{c}$. 

To understand the physical origin of these qualitative differences in the molecular association with changing intensity a simplified model is helpful.
Employing the rotating-wave approximation in Eq.(\ref{eq:3}), where the terms $\Gamma_{bb}(t)$ and $\Gamma_{\epsilon \epsilon '}(t)$ are neglected, yields a set of time-dependent equations that can be decoupled after we neglect high-order terms of the form $\partial_t^n[\chi(t) e^{iE_bt/\hbar}C_b(t)]$ with $n\ge1$.
Then the transition probability density to the molecular bound state from an initial continuum state with energy $E_\mathrm{in}$ reads

\begin{subequations}
\label{eq:5}
\begin{equation}
	|C_b(t)|^2=  \frac{\eta_0^2 |W_{b,E_\mathrm{in}}|^2}{4\hbar^2} \bigg|\int^t dt' \chi(t')e^{-\frac{\gamma J(t,t')}{2\hbar}+\frac{i}{\hbar}\Phi(t') } \bigg|^2
	\label{eq:5a}\end{equation}
with
\begin{equation}
\Phi(t)  	=\int^{t}_0\!\! dt'(E_R^*(t')-E_\mathrm{in}),\,\,\,E_\mathrm{R}^*(t)=E_b+\hbar\omega+\frac{\eta_0^2}{4}\Delta \chi^2(t)\,.
\label{eq:5b}
\end{equation}
\end{subequations}

Here $W_{b,E_\mathrm{in}}=\Gamma_{b,E_\mathrm{in}}(t)/\eta_0\cos(\omega t)$ is the bare coupling matrix element of the bound state with the initial energy-normalized continuum state and $J(t,t')=\int^t_{t'} dt'' \chi^2(t'')$.
$\gamma=0.5\pi|W_{b,E_b+\hbar\omega}|^2 \eta_0^2$ denotes the decay of the bound state into the continuum at energy $E_\mathrm{in}=E_b+\hbar\omega$ with $W_{b,{E_b+\hbar\omega}}=\Gamma_{b,{E_b+\hbar\omega}}(t)/\eta_0\cos(\omega t)$.
$\Phi(t)$ is the phase accumulated between the continuum state with energy $E_\mathrm{in}$ and the energy  $E_\mathrm{R}^*(t)$ of the shifted dressed bound state.
The shift $\frac{\eta_0^2}{4}\Delta$ is given by $\frac{\eta_0^2}{4}\Delta=\frac{\eta_0^2}{4}\mathcal{P}\int dE \frac{|W_{bE}|^2}{E-E_b-\hbar\omega}$ (the symbol $\mathcal{P} \int$ indicates the principal value integral) and depends quadratically on the strength of pulse.
Consequently, $E_\mathrm{R}^*(t)$ evolves with the square of the pulse envelope analogous to the AC-Stark shift \cite{baghery2017prl}. 

\begin{figure}[t]
\centering
 \includegraphics[scale=0.44]{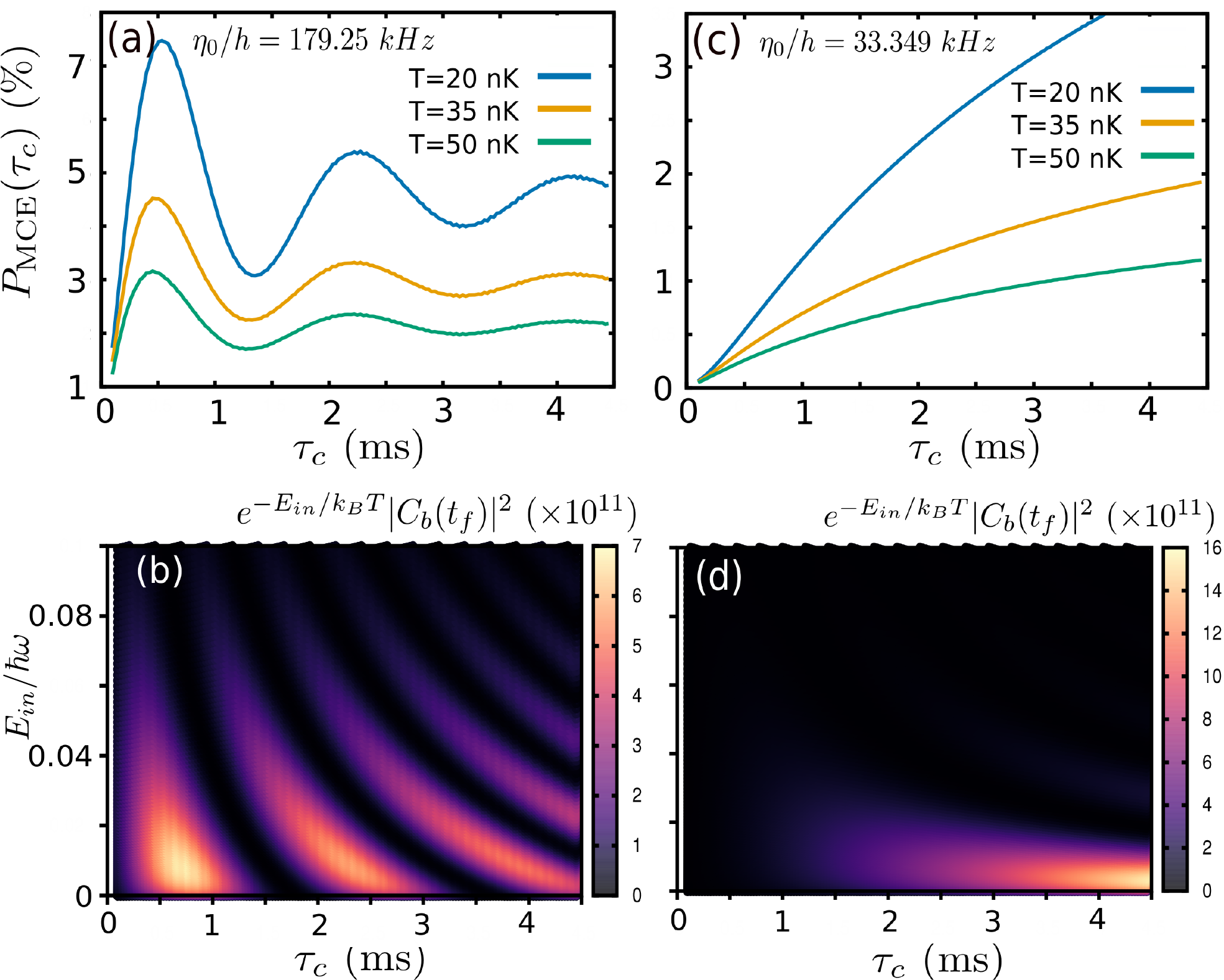}
 \caption{(Color online)  (a) and (c) The molecule conversion fraction obtained by Eqs.(\ref{eq:4}) and (\ref{eq:5}) where the strength of the pulse, the temperatures and the density of the thermal gas are the same as in Fig.\ref{fig1}(c) and (d), respectively.
	 (b) and (d) depict the integrand of Eq.(\ref{eq:4}) in a.u. at $T=20~\rm{nK}$ where pulse's strength is $\eta_0/h=179.25\,$kHz  and $\eta_0/h=33.349\,$kHz , respectively.
 }
\label{fig2}
\end{figure}

In order to gain physical insight, the time evolution of  $E_\mathrm{R}^*(t)$ in the diabatic picture is illustrated in Fig.\ref{fig1}(b) whereas the continuum is denoted by the gray shaded area.
Note that at the beginning and at the end of the pulse $E_\mathrm{R}^*(t)$ becomes equal to the bare dressed bound energy $E_b+\hbar\omega$ with $\Delta = 0$.
The blue dashed line in Fig.\ref{fig1}(b) denotes the energy $E_\mathrm{in}$ of a selected initial continuum state.
As just discussed, the dressed  bound state energy $E_\mathrm{R}^*(t)$ follows the pulse intensity envelope, and for a sufficiently strong pulse this dressed state can cross the dissociation threshold, as is the case in Fig.\ref{fig1}(c).
However, for weak pulses, cf.\ Fig.\ref{fig1}(d), $E_\mathrm{R}^*(t)$ remains in the continuum throughout the pulse.
This behavior can be intuitively understood in the context of level repulsion where during the pulse the continuum states push the dressed bound state to lower energies; hence for a strong pulse this repulsion yields an intensity-shifted, dressed bound state energy that moves below the threshold.
In addition, Fig.\ref{fig1}(b) demonstrates signatures of {\it St\"uckelberg} physics in the system: during the turn on/off of the pulse non-adiabatic couplings are induced between the dressed bound  and continuum state with energy $E_\mathrm{in}$, which translates into a two-pathway interference.
In Fig.\ref{fig1}(b) the purple dashed-dotted (green dotted) line denotes the path via the shifted dressed bound ($E_\mathrm{in}$) state.
This two-pathway interference yields the phase difference  Eq.~\eqref{eq:5b} in the wavefunction. 
The emerging picture is analogous to electron dynamics in ionization processes described with pulse envelope dependent wavefunctions \cite{tosa+15} presented in  \cite{baghery2017prl,ning2018prl} where the photoelectron experiences two non-adiabatic ionization bursts during the turn on/off of an ultrashort pulse.
The derivatives of the pulse envelope induce a two-pathway interference in time resulting in St\"uckelberg oscillations of the differential ionization cross-section as a function of the pulse length at a given electron energy.

The St\"uckelberg phase in the probability density $|C_b(t_\mathrm{f})|^2$ plays a crucial role in modulating the MCE.
More specifically, in Eq.(\ref{eq:4}) $|C_b(t_\mathrm{f})|^2$ is averaged over a MB distribution of initial continuum energies. Thereby, the corresponding St\"uckelberg phases are mixed incoherently suppressing interference fringes in the MCE.
Indeed, this behavior is observed in the Fig.\ref{fig1}(d) where the strength of the pulse is weak and throughout the pulse $E_\mathrm{R}^*(t)$ lies within the continuum.
In contrast, for strong pulses [see Fig.\ref{fig1}(c)], where the shifted dressed bound state can cross the threshold,  the St\"uckelberg phase between $E_\mathrm{R}^*(t)$ and  zero continuum energy,i.e. $\epsilon = 0$,  survives the thermal average and yields a low frequency oscillation in the molecule MCE, as we will show.

Figs.\ref{fig2} (a) and (c) show the MCE obtained by Eqs.(\ref{eq:4}) and (\ref{eq:5}) for a strong ($\eta_0/h=179.25\,$kHz) and weak ($\eta_0/h=33.349\,$kHz) pulse, respectively.
Note that the range of temperatures, turn on/off time $\tau_{0}$, driving frequency and density are the same as in Figs.\ref{fig1}(c) and (d).
In addition, for the shift $\frac{\eta_0^2}{4}\Delta$ used in Eq.(\ref{eq:5}) the Principal value integral is evaluated over the same range of continuum energies that is used in the corresponding numerical calculations.
The MCEs shown in Figs.\ref{fig2}(a) and (c) exhibit the same qualitative differences as discussed for Figs.\ref{fig1}(c) and (d).
More specifically, the MCE for strong pulses in Fig.\ref{fig2}(a) saturates at a smaller pulse length than in Fig.\ref{fig2}(c).
According to Eq.\eqref{eq:5}, this occurs due to the decay $\gamma$ which is proportional to $\eta_0^2$. Hence, a strong pulse renders the saturation of the MCE faster than a weak pulses.
Also, Fig.\ref{fig2}(a) exhibits only the low-frequency oscillations attributed to St\"uckelberg physics whereas the high-frequency ones [see Fig.\ref{fig1}(c)] are absent.
This implies that the physical origin of high frequency interference fringes is associated with higher order photon processes since in Fig.\ref{fig2}(a) the rotating-wave approximation is employed where counter rotating-wave and the $\Gamma_{bb}(t)$ terms as well as $n>1$-photon processes are neglected.
Note that due to the rotating-wave approximation the magnitude of the MCE in Figs.\ref{fig2}(a) and (c) is higher than in the corresponding Figs.\ref{fig1}(c) and (d).
Finally, Fig.\ref{fig2}(b) and (d) illustrate the integrand of MCE in a.u. [see Eq.(\ref{eq:4})] at $T=20\,$nK  as a function of the pulse's duration and the scaled energy of the initial continuum states, i.e. $E_\mathrm{in}/\hbar\omega$.
For strong pulses Fig.\ref{fig2}(b) demonstrates the interference fringes due to the St\"uckelberg phases between different initial continuum states and the shifted dressed bound one where the MB factors are considered.
In contrast, for weak pulses, see Fig.\ref{fig2}(d), the MB weights suppress the interference features of the probability density $|C_b(t_\mathrm{f})|^2$  yielding a coherence-free MCE.
Panels (b, d) demonstrate that in the MCE only continuum states with energies $E_{in}=[0,~0.1\hbar \omega]$ contribute.

To address the frequency of the slow oscillations in Fig.\ref{fig1}(b) and its dependence on temperature, Eq.(\ref{eq:5}) can be further simplified by considering a square pulse with the same FWHM as in Eq.(\ref{eq:2}).
In addition, in the temperature range of interest the coupling $|W_{b,E_\mathrm{in}}|^2$ in Eq.(\ref{eq:5}) is approximated by Wigner's threshold law, $|W_{b,E_\mathrm{in}}|^2\approx |W_{b,0}|^2 \sqrt{E_\mathrm{in}}$ where $W_{b,0}$ is the coupling of the bound state with the zero-energy continuum state.
Under these considerations, one arrives at an analytical expression of MCE in terms of the phase $\phi_{\tau\beta}=\tau_\mathrm{c}^\mathrm{eff}/\hbar\beta$ and the energy ratio $\rho_{\beta\varepsilon}=|\varepsilon_R^*|\beta$,
\begin{figure}[t!]
\centering
 \includegraphics[scale=0.48]{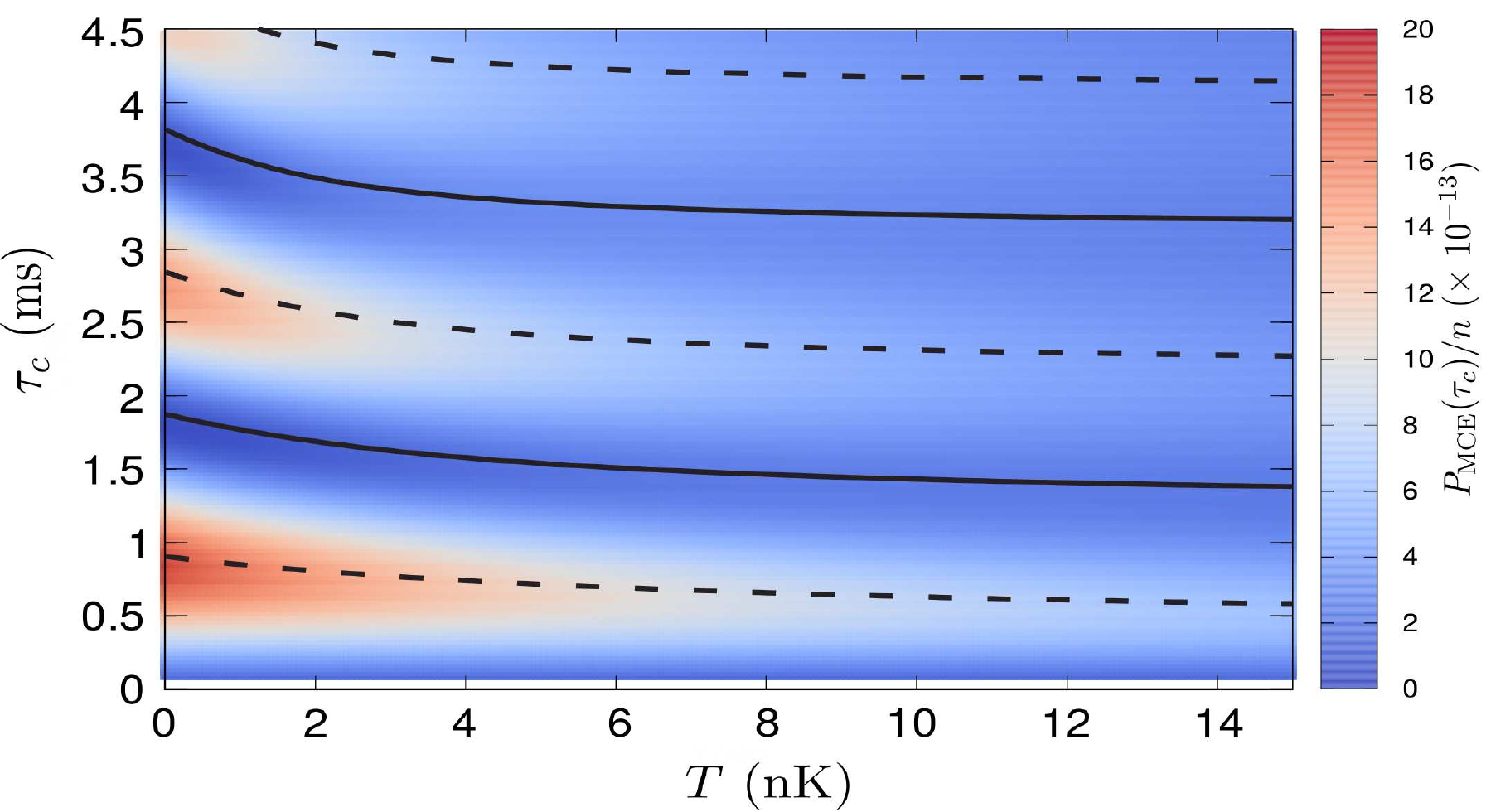}
 \caption{(Color online) $P_{\rm{MCE}}(\tau_c)/n$ as a function of the temperature $T$ and the pulse duration $\tau_\mathrm{c}$. The black dashed (solid) lines depict the maximum (minimum) of the MCE obtained by Eq.(6).
 }
\label{fig3}
\end{figure}
\begin{subequations}
\begin{align}
& P_\mathrm{MCE}(\tau_\mathrm{c}^\mathrm{eff}) =\frac{ n\lambda_T^3\eta_0^2 |W_{b,0}|^2\sqrt{\pi\beta}}{\rho_{\beta\varepsilon}(3+2\rho_{\beta\varepsilon})}	(1-g \cos \phi),\\
&  \phi =\phi_{\tau\beta}\rho_{\beta\varepsilon}+\arctan[\phi_{\tau\beta}]/2+\arctan[\frac{2\phi_{\tau\beta}\rho_{\beta\varepsilon}}{3+2\rho_{\beta\varepsilon}}]\,,
		\label{eq:6}
\end{align}		
\end{subequations}
where $\beta=1/k_B T$, $\tau_\mathrm{c}^\mathrm{eff}=\tau_\mathrm{c}+\tau_0$, $\varepsilon_R^*=E_b+\hbar\omega+\Delta \frac{\eta_0^2}{4}$ and $g=(1+\phi_{\tau\beta}^2)^{-\frac{1}{4}}[1+(\frac{2\phi_{\tau\beta}\rho_{\beta\varepsilon}}{3+2\rho_{\beta\varepsilon}})^2]^{-\frac 12}$.

Evidently, Eq.(\ref{eq:6}) captures the origin of the slow oscillations in the MCE which emerge from the thermally averaged St\"uckelberg phase $\phi=\phi(\tau_\mathrm{c}^\mathrm{eff},\varepsilon_R^*,T)$. 
The  solid (dashed) lines in Fig.\ref{fig3} show  minima (maxima) of the MCE as predicted by Eq.(\ref{eq:6}), with the full numerically obtained $P_{\rm{MCE}}(\tau_c)/n$ in the background.
One sees that the extrema approach constant values at high temperatures, i.e. at $T\gg T_s$ with $T_s=\hbar/\tau^\mathrm{eff}_ck_B$, corresponding to an asymptotic frequency  $\nu_{T\gg T_s}\approx|\varepsilon_R^*|/2\pi$ according to Eq.(\ref{eq:6}).
Consequently, the high-$T$ oscillations are due to the St\"uckelberg phase between the intensity-shifted dressed bound state and {\it the zero energy} continuum state.
The minimum pulse strength which yields St\"uckelberg interference is $\eta _0> 2\sqrt{-(E_b+\hbar\omega)/\Delta}$.
For the  parameters of Fig.\ref{fig1}(c) we obtain  $\nu_{T\gg T_s}\approx 0.515\,$kHz  which agrees with the $\nu=0.532\,$kHz  of Fig.\ref{fig1}(c).

However, in the low-$T$ regime Eq.(\ref{eq:6}) gives $\nu_{T\ll T_s}\approx(|\varepsilon_R^*|+k_BT/2)/2\pi$.
In this limit, the MB distribution is so narrow that only the continuum state with the most probable energy,  $E_\mathrm{in}=k_B T/2$, contributes in the MCE.
Therefore, $\nu_{T\ll T_s}$ depends only on the St\"uckelberg phase between the {\it most probable} continuum state energy  and $\varepsilon_R^*$.
Finally, $g$ in Eq. (\ref{eq:6})  controls the contrast in the oscillations.
For high-$T$  it behaves as $g_{T\gg T_s}\approx3\sqrt{\hbar/(k_B T \tau_\mathrm{c}^\mathrm{eff})}$ demonstrating that the contrast decreases for increasing $T$ as apparent from Fig.\ref{fig1}(c).

{\it Conclusions. \textendash} For pulsed RF association of Feshbach molecules in a thermal gas of $^{85}$Rb atoms we have demonstrated non-adiabatic effects of the pulse envelope so far only known from ionization  with ultrashort pulses \cite{ning2018prl}.
More specifically, the MCE shows interference fringes for strong RF pulses as a function of pulse length that can survive the incoherent thermal averaging in contrast to previous studies \cite{hanna2007pra}.
We have worked out the dependencies of these St\"uckelberg oscillations  on the temperature of the thermal gas  and the pulse length. In the limit of high temperatures, the oscillation frequency depends only on the St\"uckelberg phase between the zero energy continuum and the energy of the intensity-shifted dressed bound state, whereas for small temperatures
the threshold energy gets replaced by the most probable (thermal) energy.
In addition, the MCE exhibits  fast oscillations with the RF field frequency.
They are associated with higher order photon processes but will be challenging to resolve them experimentally. 

The St\"uckelberg oscillations, however, should be observable in a thermal gas of $^{85}$Rb  atoms of density $n=10^{11}/$cm$^{3}$ with a Feshbach field around $B=156.9\,$G producing the same scattering length as in our study.
The RF-pulse which is used in our calculations can be experimentally implemented by an additional  magnetic field of strength $B_m\approx0.57\,$G  modulated with  $\omega=2 \pi\times 10.11\,$kHz.
A promising direction where dynamical interferences are crucial is in molecular photoassociation processes driven by chirped short-pulses \cite{mur-petit_dynamical_2007}.
Extending the present concept into such systems including the continuum may yield non-trivial effects as was shown in the strong-field physics where chirped pulses can control the ionization \cite{saalmann_adiabatic_2018}.

{\it Acknowledgments - } This work has been supported in part by the U.S. National Science Foundation grant No. PHY-1607180, the Israel Science Foundation (Grant No. 1340/16) and the United States-Israel Binational Science Foundation (BSF, Grant No. 2012504). The numerical calculations have been performed using NSF XSEDE Resource Allocation No. TG-PHY150003. 
\bibliography{bib_assoc.bib}
\end{document}